# Electron-Spin-Resonance in a proximity-coupled MoS2/Graphene van-der-Waals heterostructure

Chithra H. Sharma,[1,a)] Pai Zhao,[1,2] Lars Tiemann,[1] Marta Prada,[3] Arti Dangwal Pandey,[4] Andreas Stierle[1,4] and Robert. H. Blick[1,5,b)]

[1] *Center of Hybrid Nanostructures (CHyN), Department of Physics, Universität Hamburg, Luruper Chaussee 149, 22761 Hamburg, Germany*

[2] *Institute of Microsystems Technology, Hamburg University of Technology, 21703 Hamburg, Germany*

[3] *Institute for Theoretical Physics, Universität Hamburg, HARBOR, Luruper Chaussee 149, 22761 Hamburg, Germany*

[4] *Centre for X-ray and Nano Science CXNS, Deutsches Elektronen-Synchrotron DESY, Hamburg, Germany*

[5] *Materials Science and Engineering, University of Wisconsin-Madison, Univ. Ave 1550, Madison, WI 53706, US*

[a)] Electronic mail: Chithra.Sharma@physik.uni-hamburg.de

[b)] Electronic mail: rblick@physnet.uni-hamburg.de

Coupling graphene's excellent electron and spin transport properties with higher spin-orbit coupling material allows tackling the hurdle of spin manipulation in graphene, due to the proximity to van-der-Waals layers. Here we use magneto transport measurements to study the electron spin resonance on a combined system of graphene and $MoS_2$ at 1.5K. The electron spin resonance measurements are performed in the frequency range of 18-33GHz, which allows us to determine the g-factor in the system. We measure average g-factor of 1.91 for our hybrid system which is a considerable shift compared to what is observed in graphene on $SiO_2$. This is a clear indication of proximity induced SOC in graphene in accordance with theoretical predictions.

Proximity causes interaction – naturally, this basic rule also holds for nano-scale junctions made of van-der-Waals (vdW) materials. [1,2] By now it is possible to assemble a large number of different vdW-materials in order to tailor a system with specific properties, thus bringing together the best of 'many worlds'. In particular, hexagonal boron nitride (h-BN) has been used to encapsulate vdW-materials, enhancing the electronic properties of graphene (Gr) as well as different transition-metal-dichalcogenides (TMDCs). [1,2] For $MoS_2$ and $WS_2$ this has been studied in p-n-junctions via excitons [3] and in twisted homo-bilayers, [4,5] which show correlated electronic phases. $MoS_2$ and Gr are heavily studied systems individually. While Gr is a semi-metal, $MoS_2$ is a semiconductor (intrinsically typically n-doped) with large spin-orbit coupling (SOC), where the carrier density and thus its conductivity can be tuned by means of a gate voltage. The mobility and conductivity of $MoS_2$, however, is limited predominantly due to the rough interface and the $SiO_2$-substrates. Also, it forms Schottky contacts with most metals, so that the fabrication of ohmic contacts is a challenge. [6] On the other hand, Gr is well-studied as a Dirac-metal with high charge carrier conductivity and mobilities. [7], but its device performance has been limited due to the lack of a band-gap and relatively small intrinsic SOC of the order of only 40μeV. [8–11] Enhancing the SOC with TMDCs in close proximity is thus a way to marry the benefits of the properties of both materials.

Graphene can be employed as an ohmic contact (or conductive backbone) material for $MoS_2$. [12,13] It is assumed that valley-Zeeman and Rashba-SOC are induced in Gr-on-TMDCs via the proximity effect. [14–18] Furthermore, it was suggested that the bandgap of $MoS_2$ could be tuned in $Gr/MoS_2$ heterostructures. [19] This induced SOC is expected to be up to two orders of magnitude higher than the intrinsic value of Gr. Concurring this, proximity induced SOC has been reported in a few experimental studies for heterostructures of $Gr/WS_2$, [20,21] $Gr/WSe_2$ [18], and $Gr/MoS_2$. [22], [18] Proximity induced spin-lifetime anisotropy has also been explored in $Gr/TMDC(MoSe_2)$. [23,24] So far, the signatures of induced SOC have been observed via magneto-transport embedded in weak anti-localization (WAL) [18,21] measurements, or via the Spin-Hall effect, [20,22] in Coulomb-drag effect studies, [25] and in the modulation of Schottky barrier height [26]. Among these only a couple of studies report the electron transport behavior in Gr-on-$MoS_2$ at low-temperatures. [18,25] A direct measurement of the *g*-factor on such a hybrid system has not been performed until today. Here the premise is that the *g*-factor for free electrons is well-known with a value of 2.0023... [27,28] Any deviation is an indication of SOC in the system. In previous resistively detected electron spin resonance (ESR)-measurements on single and multi-layer Gr, a *g*-factor, parallel to the c-axis of (1.952 +/- 0.002) was determined for Gr-on-SiO$_2$. [29]

Here, we report on a detailed resistively detected ESR traced in magneto-transport of Gr-on-MoS$_2$ (or in short GroMoS) samples in direct comparison with standard Gr-on-SiO$_2$ (or GroSi). The four-point measurements of the electrical resistance of the device is measured at 1.5K, as shown in the schematic architecture in Fig. 1(a). In order to couple an ESR-signal, a Hertzian resonator coil is placed in vicinity to the sample for microwave irradiation which will be discussed in the following. [29] The magnetic field component of the microwave ($\boldsymbol{B_\nu}$) has to be perpendicular to the external magnetic field ($\boldsymbol{B}$). [30] The magnetic moment of the electron precesses around the direction of $\boldsymbol{B}$ while a resonant $\boldsymbol{B_\nu}$ tips the magnetic moment into the plane i.e., perpendicular direction to $\boldsymbol{B}$. In terms of energy, the spin energy level splits with $\boldsymbol{B}$, while the $\boldsymbol{B_\nu}$ causes the spin flips when the frequency matches the splitting, according to the equation $h\nu = g\mu_B B$. These resonant spin flips are observed as a change in the resistance attributed to spin energy being transferred to mobile electrons and momentum randomization of electrons. [31–33]

We use standard magneto-transport measurements to detect ESR, evaluate the charge transfer between the layers, and determine the doping density. We are able to report on the deviation of the *g*-factor (*g* // c-axis) of Gr from its intrinsic value in the heterostructure signaling that the SOC is altered by the close proximity of the MoS$_2$-layer.

We exfoliated few-layer MoS$_2$ on a 300nm SiO$_2$/Si wafer and fabricated a Hall bar of CVD grown Gr [Graphenea Semiconductor S.L., Spain] on top. [29,34] A part of the Gr-Hall-bar covers the MoS$_2$ sample, while the other part is directly placed on SiO$_2$, as shown in Fig.1 (b). The contacts were fabricated on top of the Gr on both regions using standard electron beam lithography techniques with Ti/Au (4nm/70nm) metallization. Fig. 1(b) shows the optical image of the device where the exfoliated $MoS_2$ flake is outlined by the purple dotted line as a guide to the eye. The reactive ion etching process for the Hall-bar fabrication (also etched the $MoS_2$ outside the defined region) ensures ohmic contacts on Gr. The highly *p*-doped Si under the 300nm thick $SiO_2$ dielectric was used as the gate electrode to tune the carrier concentration.

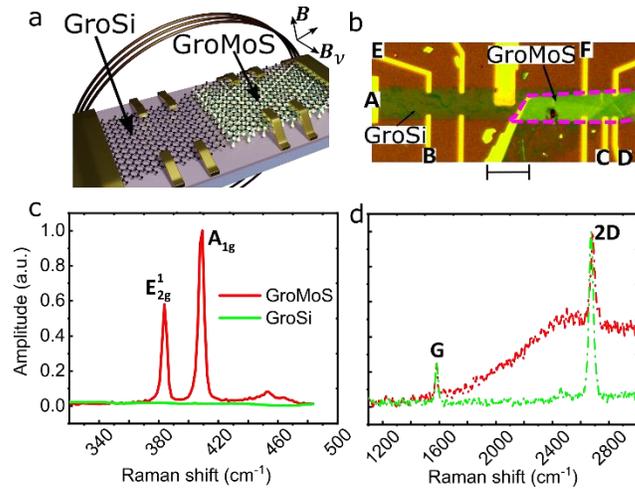

**Fig. 1** (a) Schematic representation of the device architecture including the Hertzian-loop antenna for coupling the microwave signal for ESR. (b) Optical image of the device: the dotted purple region traces the outline of the $MoS_2$-flake. The contacts used for the measurements are labeled as A-F. The scale bar is 20μm. (c, d) The Raman spectra taken from the GroMoS and GroSi regions are shown in red and green, respectively. (c) $E^1_{2g}$ and $A_{1g}$ peaks of $MoS_2$ (d) G and 2D peaks of Gr as dotted lines (see text for details).

In order to characterize GroSi and GroMoS we performed Raman spectroscopy with a standard 532nm laser, as shown in Fig. 1(c) and (d). Fig. 1(c) shows the peaks corresponding to $MoS_2$ where the in-plane ($E^1_{2g}$) and out-of-plane vibration peaks ($A_{1g}$) are observed at 383.8cm$^{-1}$ and 408.6cm$^{-1}$, respectively only on the GroMoS. Clear G and 2D peaks of Gr are observed in both the regions in Fig. 1(d). We do not observe any significant D-peak, indicating that the quality of Gr has not been compromised. We observe that GroMoS shows a small upshift in the 2D-peak position compared to GroSi. The observed upshift is consistent with the vdW-interaction with the $MoS_2$ flake [35]. Atomic force microscopy (AFM) performed on the sample after the transport measurements reveals that uniformity of the graphene layer on MoS$_2$ has not been compromised (see Supplementary Material Figure S1). The thickness of the flakes measured by AFM is higher than attributed to the moisture adsorption and thermal cycling.

The device was annealed after fabrication at 200°C in vacuum of 10$^{-4}$ mbar overnight to remove any excess moisture. [34] The sample was then cooled to 1.5K in a helium bath cryostat without breaking the vacuum. All transport measurements shown are performed at 1.5K with and without a perpendicular magnetic field. The results to be discussed further are measured in four-point lock-in configuration by passing an AC current across the probes marked as A and D (ground) of Fig. 1(a).

Figure 2 (a) shows the longitudinal resistance of the device measured between the probes B and C as a function of the applied back gate-voltage ($V_g$). This measurement was performed over the whole Gr Hall-bar where one voltage probe is on GroSi and the other on GroMoS. The black and the green dotted-line traces show the forward and backward sweep respectively, where a clear ambipolar behavior is observed, as expected in Gr. The solid traces are respective Lorentzian fits to the measured trans-resistance. The charge neutrality point (CNP) for the Gr is $V_g \sim 7$V. The doping of the graphene is appreciably small and we do not observe a significant hysteresis, which would be indicative of the reduced moisture on the sample. [34]

As observed commonly for CVD-grown Gr, we find weak localization (WL) signatures as a result of the strong inter-valley scattering. [36] WL is observed for all gate-voltages with the data close to the CNP shown in Fig. 2(b). The Hall resistance measured across the probes B and E for different gate voltages are shown in Fig. 2(c). Note that this Hall measurement is performed in the GroSi-region alone. The sign reversal of the slope between 7-8V signals the change in the carrier type from holes to electrons at the CNP. We also observe a small WL-peak in the Hall data from a small parasitic longitudinal component in the measurement. The Hall slope is plotted against the gate-voltage in Fig. 2(d), where the sign change is most evident. The mobility of the Gr is estimated to be ~500cm$^2$V$^{-1}$s$^{-1}$. The poor mobility in the device could be a result of the grain boundaries and wrinkles in our samples as observed in the AFM image (Supplementary Material Figure S1). [37]

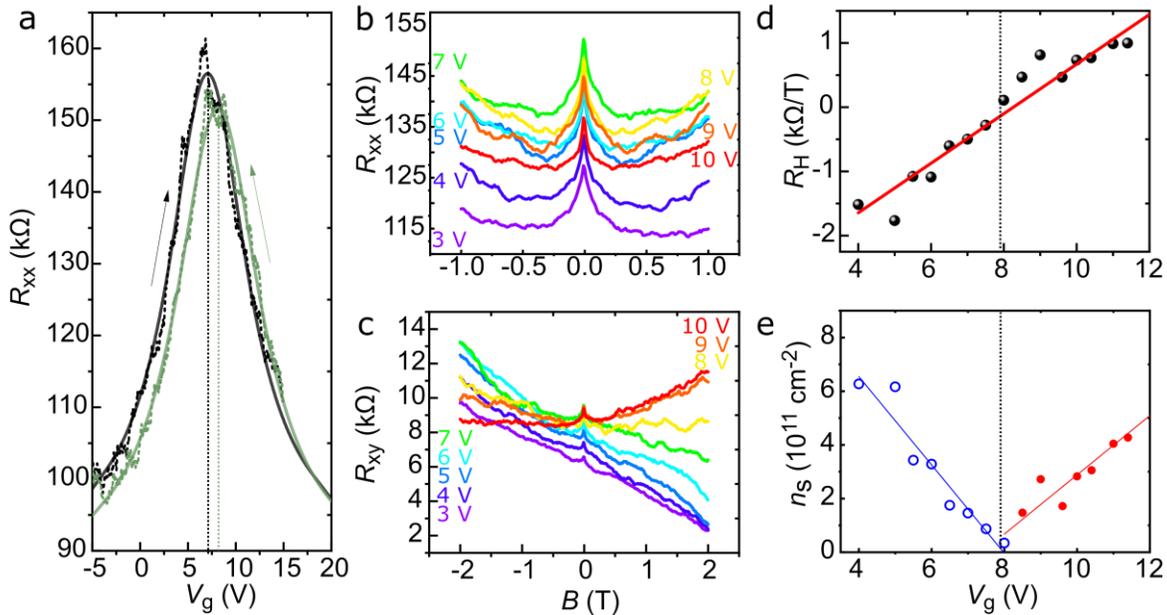

**Fig. 2** (a) Trans-resistance of the device measured across B and C, which is sampling over the Gr covering SiO$_2$ as well as the MoS$_2$. The black and green traces show forward and backward sweeps, respectively, in dotted lines, while the full lines we show Lorentzian fits to the data. (b, c) Magneto-transport of the device at different gate voltages measured between contacts (b) B and C showing weak localization and (c) B and E showing the Hall resistance. The gate voltages are indicated by the color scale in the graphs. (d) Variation of the Hall slope as a function of gate voltage close to the CNP indicating that we are dealing with a hole-carrier system in total. (e) Carrier density as a function of gate voltage close to the CNP calculated based on the two-carrier model. [See Supplementary Material S2 for details]

The carrier density is calculated close to the CNP using a two-carrier model [38] (for details see Supporting Material S2) and plotted in Fig. 2 (e) using the longitudinal and transverse resistances in

Fig. 2 (b) and (c) respectively. The blue open circles and the red filled circles represent the holes and electrons, respectively. The blue and red lines are linear fits to the carrier density that is lowest around ~ 8V. We also note a small asymmetry in the change of carrier density with the gate-voltage for the hole and the electron side. We attribute this to the GroMoS region in the device. A saturation of the electron carrier density has been observed previously in GroMoS devices as a result of the negative compressibility in the system. [39]

From a separate measurement on the GroMoS region (see Supplementary Material Figure S2) we observe that the CNP is shifted to ~ -6V, implying that MoS$_2$ is *n*-doping the Gr and thus a strong interaction between the systems is induced. We also observe jumps in the *I-V*-characteristics taken from the GroMoS region as opposed to the whole sample, which is indicative an exchange of electrons between the two layers (Supplementary Material Figure S3). We do not observe any ESR from this region, presumably as the cross section of the probed region is too small to induce a measurable signal.

Proximity induced SOC interactions between Gr and MoS$_2$ is reflected in the electronic *g*-factor that we can experimentally address via ESR. Here, the magneto-resistance is measured across contacts B and C, while simultaneously applying microwave radiation using a Hertzian coil antenna, as shown in the schematic in Fig. 1(a). The resonance signal is observed in the difference Δ$R_{xx}$ of the magneto-transport measurements with and without radiation (dark), $\Delta R_{xx}(B, h\nu) = R_{xx}(B, h\nu) - R_{xx}(B, dark)$. The signal peak follows the resonance condition $h\nu = g\mu_B B$ corresponding to the Zeeman splitting for various microwave frequencies $\nu$. [40]

Figure 3(a) shows exemplary traces of Δ$R_{xx}$, measured at a gate voltage of 6.1V, with ESR peaks highlighted by a dotted line as a guide to the eye. The smaller non-dispersive peak centered at *B* = 0 is an artefact from the temperature-dependence of WL under non-resonant heating. The dense color-scale plot in Fig. 3(b) is a summary of all frequencies. The resonance region is marked with magenta dashed-line. Fig. 3(c) shows a linear fit of the resonance, where the slope reflects the g-factor at 6.1V. Similarly, we are now able to extract the *g*-factor for all other gate voltages around the CNP. The result of this *g*-factor analysis is plotted in Fig. 3(d) over the gate voltage range of interest. A detailed inspection results in an average value of 1.91 (dashed red line) as opposed to earlier reports on pure Gr, which reported a constant value of 1.952 +/- 0.002, irrespective of the gate-voltage. [10,29] We also calculated the spin life-time using $\tau_s = \hbar/2\Delta E \cong (71.5 \pm 4)$ps where $\Delta E = \mu_B \Delta B$ is the Zeeman energy and $\Delta B$ is the half width of the ESR peaks. [29] This is comparable to reported out-of-plane spin lifetime values for graphene and Gr/TMDC. [23,29] The half-width (spin life-time) is slightly smaller (larger) in comparison to that measured in GroSi reported by Lyon *et al*. [29] The graphene flake in our device lies partly over $MoS_2$. We believe this lowers the electron-hole puddles caused by the $SiO_2$ substrate thereby increasing the spin life-time. [41]

In contrast to former measurements, we observe a strong variation of the *g*-factor with gate voltage (i.e., carrier concentration). For positive $V_g$, the *g*-factor correction is twice as large as that for pure Gr, [29] which is consistent with the SOC enhancement via the proximity effect. As described by Gmitra *et al.*, [16] a positive electric field induces a carrier transfer between graphene and $MoS_2$ where first principal calculations estimate a splitting of ~ 1meV. Hence, we can assume that the *g*-factor variation from 1.95 to 1.91 and its dependence on the gate voltage is proximity-induced, signaling the interaction with the $MoS_2$-layers. We stress that the measurement was carried out over a large strip of Gr covering $SiO_2$ as well as $MoS_2$. This 'mixed' approach was necessary to improve the signal-to-noise ratio in $\Delta R_{xx}$. We have not been able to obtain a recognizable ESR signal when the distance between the voltage probes is reduced. This, we assume is due to the smaller number of spin flips owing to smaller area.

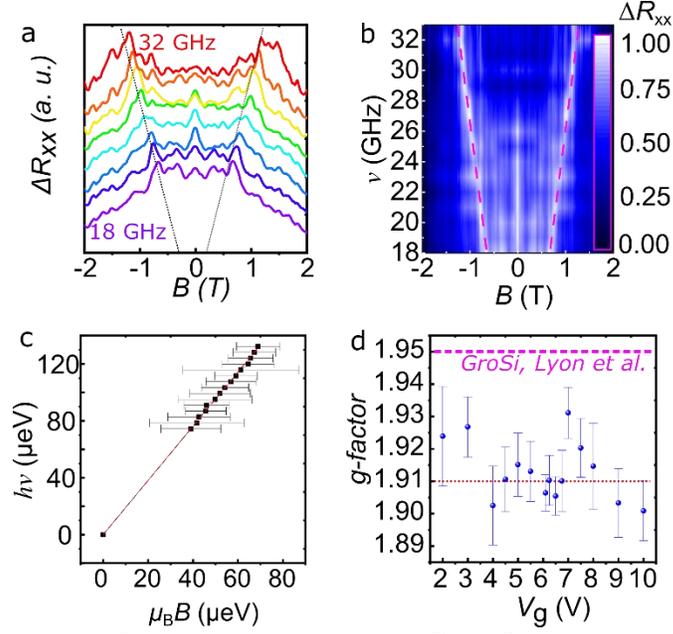

**Fig. 3** (a) Magneto-transport of the device between contacts B and C under microwave radiation at $V_g = 6.1$ V. Plotted is $\Delta R_{xx}$, as the difference in magneto-resistance with and without continuous microwave radiation. The ESR-signal is clearly discernible, dispersing linearly (dotted lines) with frequency from 18 to 32GHz. (b) Color-scale plot of the detailed frequency dependence. The plots are normalized for clarity. The resonance is marked with magenta dashed line (c) Analysis of all ESR-data for $V_g = 6.1$V revealing spin resonance (fits are anchored at zero). (d) Extracted $g$-factor of all ESR-data over the gate voltage range of interest next to the CNP. The average value is marked with the red dashed line, the measured value for Graphene on $SiO_2$ from [29] is marked in magenta dashed line.  (see text for details).

In conclusion, we can state that a clear ESR-signal in heterostructure of Gr/$MoS_2$ is traceable, due to the interaction between the two systems. Variations of the $g$-factor indicate cross-talk of the strong spin-orbit-coupled electrons of $MoS_2$ to the carriers in Gr as expected from theoretical calculations [14–18].  Enhanced SOC in Gr can potentially lead to topological phases such as quantum spin Hall effect. [11]  Understanding the exact nature of this coupling will enable us to design spin-transfer devices with possibly extremely enhanced spin-relaxation times.

## SUPPLEMENTARY MATERIAL

See Supplementary Material for AFM image of the device, Description of two-carrier model for graphene, Trans-resistance measurement in GroMoS region and *I-V*-characteristics.

## ACKNOWLEDGEMENT

This work has been supported in the early stages by the Center for Excellence Advanced Imaging of Matter (AIM) of the Deutsche Forschungsgemeinschaft (DFG) under grant number EXC-2056 and Partnership for Innovation Education and Research (PIER) under PIF-2021-01. We are grateful for direct support by the Alexander von Humboldt foundation for C.H. Sharma. We thank Prof. H.P. Oepen for discussions on spin-transport. We thank Jann Harberts for the help and suggestions on the schematics. All measurements in this work were performed with NANOMEAS. RHB likes to thank Alberto F. Morpurgo for enlightening discussions on the prospects of vdW-materials.

## AUTHOR DECLARATIONS

**Conflict of Interest**

The authors have no conflicts to disclose.

## DATA AVAILABILITY

The data that support the findings of this study are available from the corresponding author upon reasonable request.